\documentclass[10pt,conference]{IEEEtran}
\IEEEoverridecommandlockouts
\usepackage{cite}
\usepackage{amsmath,amssymb,amsfonts}
\usepackage{algorithmic}
\usepackage{graphicx}
\usepackage{textcomp}
\usepackage{xcolor}
\usepackage{algorithmic}
\usepackage{algorithm}
\usepackage{url}
\usepackage{amsmath} 
\usepackage{adjustbox}
\usepackage[misc]{ifsym}
\usepackage{booktabs}

\def\BibTeX{{\rm B\kern-.05em{\sc i\kern-.025em b}\kern-.08em
    T\kern-.1667em\lower.7ex\hbox{E}\kern-.125emX}}

\begin{document}

\title{MegaCacheX: Towards Cost-Effective Hierarchical Collaborative Content Caching in Emerging Mega-Constellations\\
}
\author{
Haoyang Shi, Xing Zhang\textsuperscript{*}, Sitong Li, Minghang Li, Xinming Lu, Shaoxiang Xu, Guoquan Wang\\
School of Information and Communication Engineering\\
Beijing University of Posts and Telecommunications, Beijing, China\\
Email: zhangx@ieee.org
}
\vspace{-10pt}

\maketitle
\begin{abstract}
Significant latency issues in global content delivery primarily arise from insufficient terrestrial network infrastructure. Deploying space-based content delivery networks within emerging mega-constellations provides an effective solution to bridging the digital divide. However, space-based caching is constrained by physical-layer dynamics, including dynamic topologies, time-varying inter-satellite link conditions, and limited onboard energy resources. Moreover, existing caching mechanisms typically lack fine-grained content categorization and comprehensive global optimization. In this paper, we proposed MegaCacheX, a cost-effective and hierarchical collaborative content distribution framework that achieves “Earth-Independence” by providing cloud services directly from space. Specifically, data centers in Sun-synchronous orbit (SSO) act as primary content sources, while caching nodes within mega-constellations and ground stations collaboratively constitute a distributed edge layer. MegaCacheX optimizes caching strategies by integrating content popularity, regional user distribution, and satellite trajectory predictions. Multi-tier caching nodes serve as service anchors, ensuring seamless content delivery with low latency. We implemented a MegaCacheX prototype on a microservices-based and containerized testbed. Evaluation results indicate that MegaCacheX reduces global content access latency by approximately 36\% compared to baseline approaches, while maintaining cost efficiency.
\end{abstract}

\section{Introduction}
The rapid development of LEO mega-constellations (e.g., SpaceX’s Starlink\cite{Starlink}) has positioned satellite-based internet services as a key technology for addressing the limitations of terrestrial infrastructure. As illustrated in Fig.~\ref{fig2}a, these systems offer ubiquitous connectivity to remote regions and advance space-air-cloud integrated architectures through deep integration with cloud computing platforms—exemplified by "Ground Station as a Service" offerings such as AWS Ground Station\cite{AWS} and Azure Orbital. This integration enables single-hop access to cloud edge computing via satellite networks, significantly reducing latency for time-sensitive applications in disaster response, healthcare, and finance. Meanwhile, Sun-synchronous orbit (SSO)-based spaceborne data centers achieve gigawatt-scale, low-carbon operations by leveraging continuous solar illumination, made possible through integrated spaceborne solar power generation and radiative cooling systems \cite{Starcloud}.
\begin{figure}[tbp]
\centerline{\includegraphics[width=\linewidth]{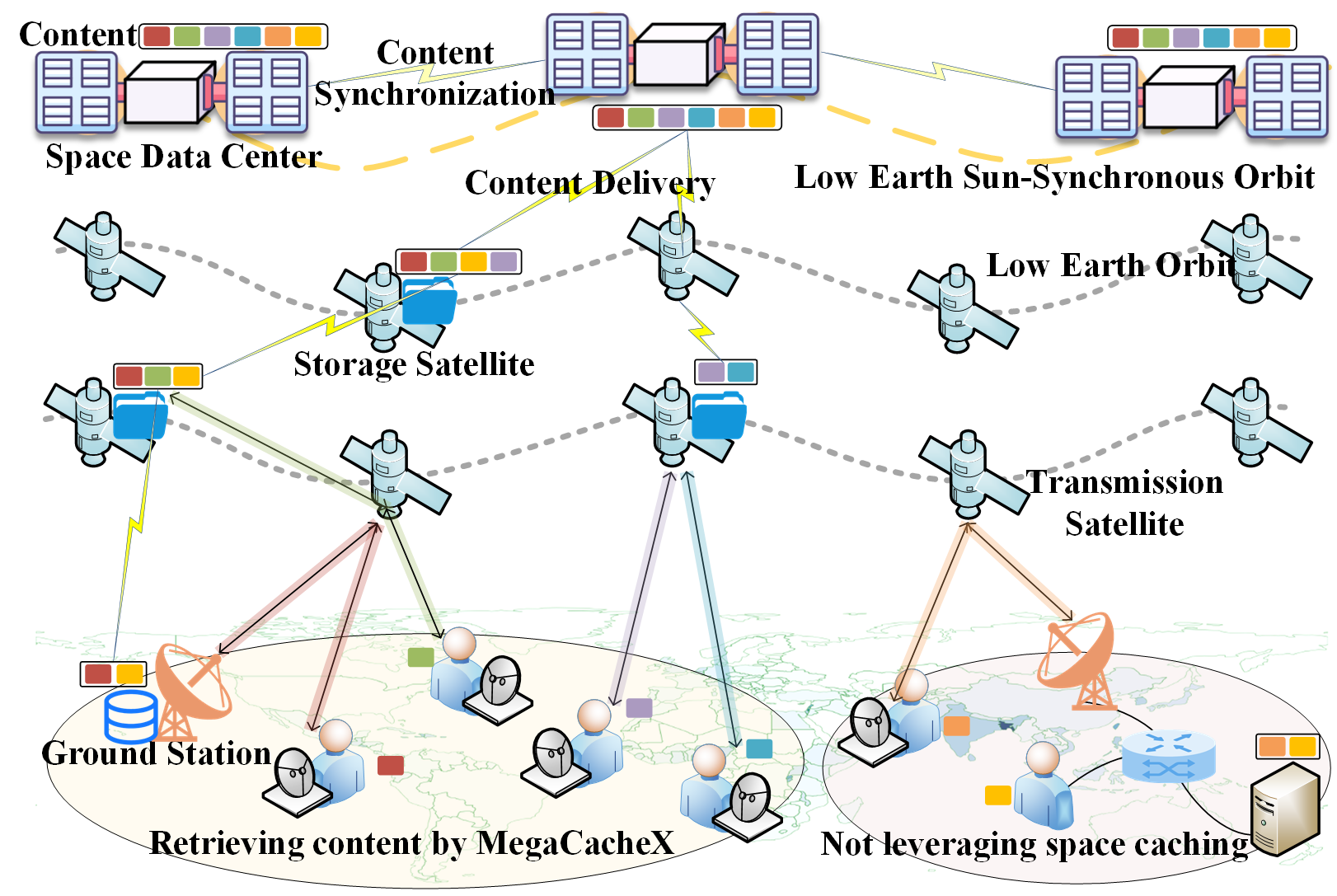}}
\vspace{-10pt}
\caption{The overview of MegaCacheX architecture.}
\vspace{-5pt}
\label{fig1}
\end{figure}

Latency-sensitive applications such as web browsing, email access, and multimedia streaming critically depend on efficient content delivery, as prolonged latency directly degrades user experience. Content Delivery Networks (CDNs) mitigate internet transmission delays by deploying distributed edge servers to enable proximity-based content caching and delivery. However, terrestrial CDNs remain constrained by their reliance on fiber-optic backbone networks and regional data centers, resulting in insufficient coverage for remote areas with underdeveloped infrastructure.
\begin{figure*}[tbp]	
    \center{\includegraphics[width=\textwidth] {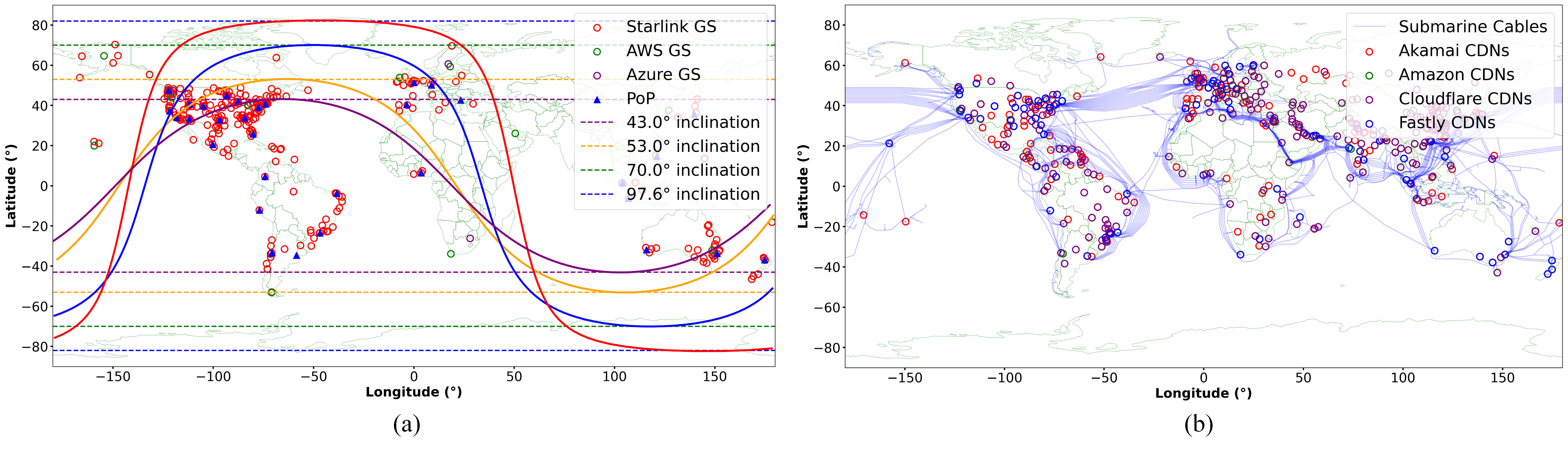}} 
        \vspace{-30pt}
    \caption{(a) Terrestrial infrastructure distribution of satellite operators. (b) Terrestrial infrastructure distribution of CDN operators.}
        \vspace{-15pt}
    \label{fig2}
\end{figure*}

\begin{figure*}[tbp]
    \center{\includegraphics[width=\textwidth] {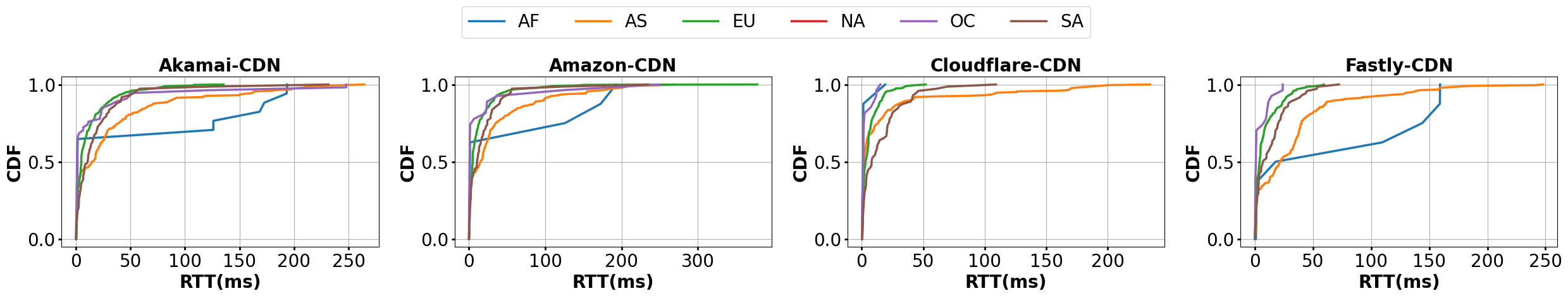}} 
        \vspace{-25pt}
    \caption{Latency distribution of international CDN operators across continents.}
        \vspace{-20pt}
    \label{fig3}
\end{figure*}

Our measurements of Round-Trip Time (RTT) across four global CDN operators (Akamai, Amazon, Cloudflare, Fastly) in 469 cities on six continents\cite{Globalping} reveal persistent latency disparities. As shown in Fig.\ref{fig2}b, despite extensive submarine cable deployments and regional CDN nodes, infrastructure investments are concentrated in economically developed regions, exacerbating the global digital divide. Fig.\ref{fig3} indicates that over 18\% of users experience RTT exceeding 50 ms, with 20\% of African users facing latency above 150 ms. Even in developed regions, approximately 8\% of users encounter delays around 100 ms due to suboptimal routing paths caused by geopolitical constraints and policy-driven routing fragmentation.

These findings underscore the inherent limitations of terrestrial CDNs and highlight the necessity for space-based CDN solutions. Mega-constellations offer the potential to bridge terrestrial coverage gaps, yet three critical challenges remain: (1) orbital storage requires a continuous energy supply, but traditional satellites struggle to support high-power caching under intermittent power availability; (2) the absence of global content planning leads to redundant replicas and cold data retention, substantially increasing storage costs; and (3) the dynamic topology of satellite networks, combined with the spatiotemporal variability of user requests, complicates global content scheduling.

Research into space-based caching leveraging orbital networks has explored satellite–terrestrial cloud coordination\cite{8}, spaceborne cloud-native control planes\cite{9}, popularity-aware caching\cite{10}, and density-aware network partitioning\cite{11}. However, most studies overlook key operational constraints, including the high energy demands of orbital storage, heterogeneous content value, and the significant capital and operational costs of spaceborne infrastructure.

To address these critical challenges, we propose MegaCacheX (where “X” denotes system extensibility), a hierarchical, multi-tier collaborative caching architecture, as illustrated in Fig.~\ref{fig1}. The framework consists of the following components: Tier-1 Cache System: Sun-synchronous orbit (SSO) data centers function as the primary source nodes of the space-based CDN (see Fig.~\ref{fig4}), supporting large-scale data storage, computing, and intelligent processing. Benefiting from stable solar illumination, these centers enable continuous 24/7 data distribution and real-time content synchronization with other orbital data centers, thereby ensuring timely updates and consistent data availability throughout the space-based CDN. 2) Tier-2 Cache System: Mega-constellation caching nodes, composed of distributed LEO satellites, provide low-latency, high-coverage content delivery and edge computing through inter-satellite collaboration and pre-caching strategies. 3) Tier-3 Cache System: Ground stations, leveraging their fixed infrastructure—such as a continuous power supply and high-capacity storage—offer extended content preservation compared to satellites. This makes them particularly suitable for caching long-tail content (i.e., data that is infrequently accessed but must remain persistently available), thereby strengthening CDN resilience. This tri-tiered design effectively addresses challenges related to latency, coverage, and infrastructure limitations, while ensuring seamless service continuity. 

This paper makes three contributions: (i) empirical evidence from global CDN measurements showing that terrestrial infrastructure constraints and regional routing fragmentation induce significant latency, motivating space-based caching; (ii) MegaCacheX, a modular, multi-tier caching architecture for mega-constellations that adapts to heterogeneous orbital deployments; and (iii) MLC3, a collaborative caching strategy that leverages hierarchical content placement to reduce latency and improve cost-efficiency across space–ground networks.

\section{SYSTEM MODEL}
\begin{figure}[htbp]
    \begin{minipage}{0.52\linewidth}
        \centerline{\includegraphics[width=\textwidth, keepaspectratio]{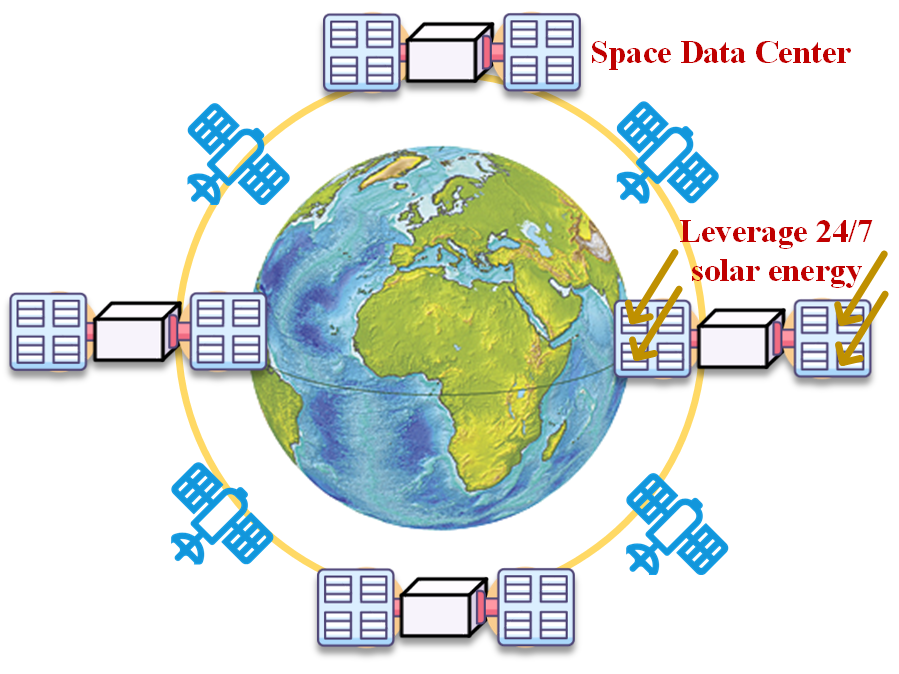}}
    \end{minipage}
    \begin{minipage}{0.47\linewidth}
        \centerline{\includegraphics[width=\textwidth]{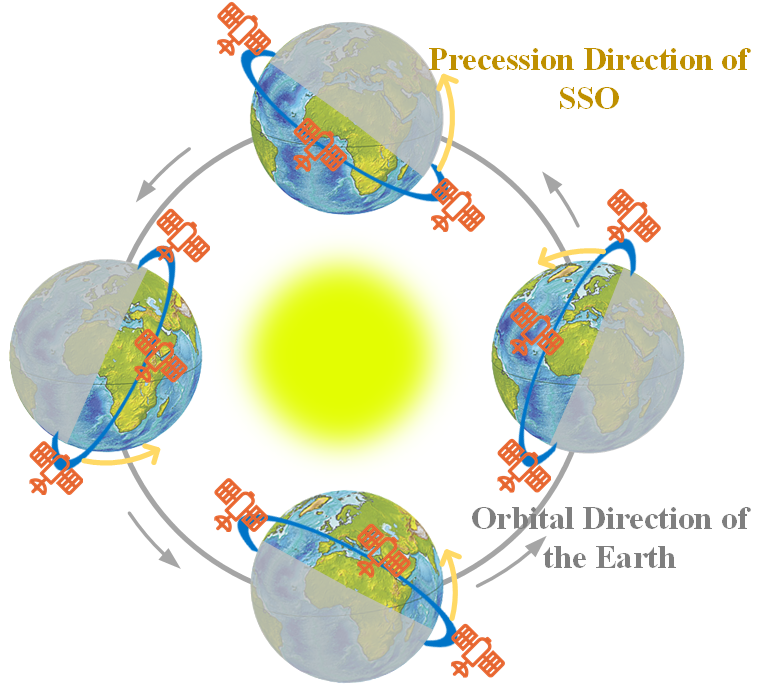}} 
    \end{minipage}
    \caption{Space-based data center Facility in Sun-Synchronous Orbit.}
    \label{fig4}
    \vspace{-5pt}
\end{figure}

\begin{figure}[tbp]
	\center{\includegraphics[width=\linewidth] {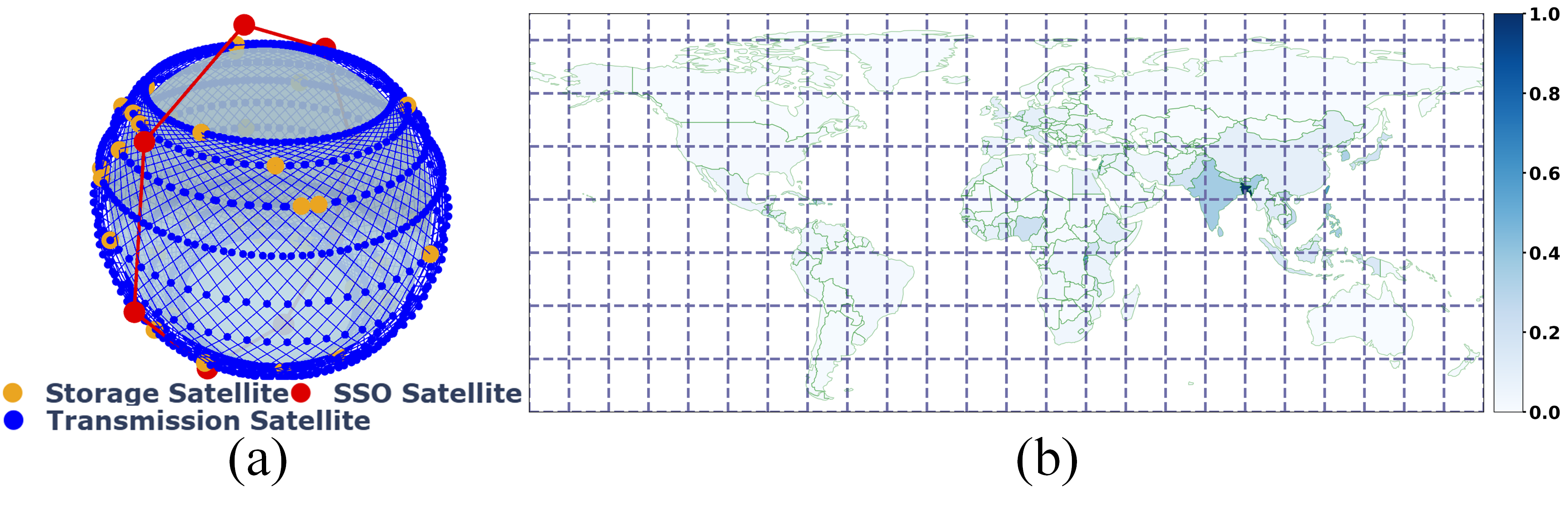}} 
        \vspace{-25pt}
	\caption{(a) Space node set \(SN\). (b) Geospatial partitioning of global service areas \(A\) and request distribution.}
	\label{fig5}
\vspace{-5pt}
\end{figure}

The mega-constellation system is modeled as a time-varying graph to capture its dynamic complexity arising from multiple functional domains. Let \(T\) represent the system’s operational period, which is divided into discrete time slots denoted as
\(T=\left \{t_{1} ,t_{2} ,t_{3} ,..., t_{\left | T \right | }\right \}\). The satellite set is defined as \(S=\left \{s_{1} ,s_{2} ,..., s_{n} , ...\right \} \). The space data center located in a sun-synchronous orbit (as shown in Fig.~\ref{fig4}) is defined as \(SDC=\left \{sdc_{1} ,sdc_{2} ,..., sdc_{i} , ...\right \} \). The union of satellites and space data centers forms the space node set \(SN=S\cup SDC\), as illustrated in Fig.~\ref{fig5}a. On the ground, stations and data centers are respectively defined as \(GS=\left \{gs_{1} ,gs_{2} ,..., gs_{m} , ...\right \}\), and \(GDC=\left \{gdc_{1} ,gdc_{2} ,..., gdc_{j} , ...\right \}\), with the ground node set given by \(GN=GS\cup GDC\). The complete node set of the system is \(V=GN\cup SN\).

Each node \(v\in V\) has a time-dependent storage capacity \(C_v^t\)during time slot \(t\). The cached content library \(F=\left \{f_{1} ,f_{2} ,..., f_{q} , ...\right \}\) contains items with size \(B_f\) for content \(f\). The global service area is partitioned into regions \(A=\left \{a_{1} ,a_{2} ,..., a_{v} , ...\right \}\), as shown in Fig.~\ref{fig5}b, based on the geographic locations of nodes or their sub-satellite points. Each node \(v\) is associated with a region \(a_v\in A\). The number of user requests for content \(f\) in region \(a\) at time \(t\) are captured by \(u_{f,a}^{t}\), with the global request distribution tensor \(U=(u^t_{f,a})_{a\in A,f\in F,t\in T}\).

\section{Multi-Level Collaborative Content Caching (MLC3) strategy for MegaCacheX}
The communication infrastructure comprises three link types: inter-satellite links (\(ISL_{(x,y)}^{t}\in \left \{0,1\right \} \)) between satellites \((x,y)\in SN\), ground-space links (\(GSL_{(x,y)}^{t}\in \left \{0,1\right \} \)) connecting ground nodes \(x\in GN\) to space nodes \(y\in SN\), and inter-ground links (\(IGL_{(x,y)}^{t}\in \left \{0,1\right \} \)) between ground nodes \((x,y)\in GN\). A link exists at time \(t\) if its corresponding binary variable equals 1. The edge set \(E^t\) aggregates all active links at \(t\), defined as \(E^t=\left \{ (x,y)|ISL_{x,y}^t \vee GSL_{x,y}^t \vee IGL_{x,y}^t=1 \right \} \). The temporal evolution of the system’s connectivity is therefore described by the graph sequence \(G=\left \{ G^t \right \} _{t\in T}\), where each snapshot \(G^t=(V, E^t)\) represents the network topology at time slot \(t\).

The data link performance in the MegaCacheX system is quantified by: 
\begin{equation}
LP^t=TBS^t \cdot R \cdot \log_{2}{M^t} \cdot f(SNR^t,TP^t)
\label{eq1}
\end{equation}
where \(TBS^t\) denotes the transport block size, \(M^t\) represents the modulation order (bits per symbol), \(R\) is the coding rate, \(SNR^t\) signifies the signal-to-noise ratio, and \(TP^t\) indicates the transmit power. The function \(f(\cdot)\) is Shannon formula. Fig.~\ref{fig8} illustrates the per-minute distributions of $LP^t$ parameters—\(TBS\), \(SNR\), \(R\), \(TP\), and \(M\)—over a 6-minute interval, derived from field measurement data\cite{NTN}.

The content delivery latency between user \(u\) and service anchor \(v\) can be modeled by:
\begin{equation}
L_{(u,v),f}^{t} =\frac{B_f}{\omega\cdot LP^t} + \frac{D_{(u,v)}}{\psi \cdot c},\forall u\in U_{F,A}^{T} ,v\in V, f\in F
\label{eq}
\end{equation}
which combines transmission delay \(\frac{B_f}{\omega\cdot LP^t}\) and propagation delay \(\frac{D_{(u,v)}}{\psi \cdot c}\). Here, \(\omega \in [0,1]\) serves as a weighting factor for link performance optimization, \(D_{(u,v)}\) denotes the communication distance between nodes, and $c$ represents the speed of light in space.

The time-dependent popularity \(p_f^t\) of content \(f\) can be derived from
\begin{equation}
p_f^t=\alpha \cdot u_{f,a}^{t} + (1-\alpha)(\sum_{i=1}^{t-1} p_f^i\cdot \theta ^{t-i}), \forall f\in F
\label{eq}
\end{equation}
integrating both real-time user requests \(u_{f,a}^{t}\) and historical popularity decay \(\sum_{i=1}^{t-1} p_f^i\cdot \theta ^{t-i}\). The weighting coefficient \(\alpha\in [0,1]\) balances immediate and historical trends, while \(\theta\) governs the exponential decay rate.

For ground station caching decisions, the probability \(P_{gs,f}^{t}\) of caching content \(f\) at station gs can be determined by
\begin{multline}
P_{gs,f}^{t} = \frac{\frac{\sum_{t-p}^{t}u_{f,a_{gs}}^{t}}{\sum_{t-q}^{t}\sum_{f\in F} u_{f,a_{gs}}^{t}} \cdot \frac{\sum_{u\in U_{f,a_{gs}}^{t}}\sum_{v\in GS}L_{(u,v),f}^{t}}{\sum_{u\in U_{f,a_{gs}}^{t}}L_{(u,gs),f}^{t}} \cdot \frac{B_f}{C_{gs}^t} }{max(P_{GS,F}^{T})}, \\
\forall gs\in GS, f\in F \label{eq}
\end{multline}

\begin{figure*}[tbp]
    \center{\includegraphics[width=\linewidth] {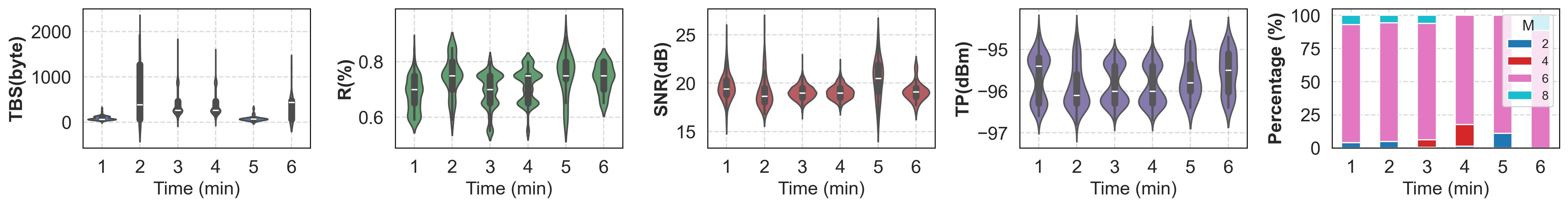}} 
    \vspace{-25pt}
    \caption{In-Field Data Link Performance Characterization: Statistical Distribution Analysis of Operational Metrics.}
    \label{fig6}
    \vspace{-20pt}
\end{figure*}

This metric aggregates two critical factors: the proportion of content-specific requests within a sliding window \(q\), and the normalized latency reduction achieved by caching \(f\) locally. By prioritizing frequently requested content in proximity to users, this mechanism significantly reduces access latency while avoiding redundant storage.
\begin{multline}
P_{s,f}^{t} =\frac{\frac{p_f^t}{\sum_{f\in F}p_f^t} \cdot \frac{\sum_{u\in U_{f,a_{s}}^{t}}\sum_{v\in S\cup SDC}L_{(u,v),f}^{t}}{\sum_{u\in U_{f,a_{s}}^{t}}L_{(u,gs),f}^{t}} \cdot \frac{B_f}{C_s^t}}{max(P_{S,F}^{T})}, \\
\forall s\in S, f\in F \label{eq5}
\end{multline}

Equation (5) governs the satellite-tier caching strategy, where popular content is systematically cached closer to edge nodes while less popular items are retained near core servers. This hierarchical approach minimizes user-perceived latency, enhances cache utilization efficiency, and provides upstream data support for in-orbit real-time computing services. The formulation explicitly accounts for spatiotemporal request patterns, link quality fluctuations, and storage capacity constraints, thereby ensuring balanced performance across the mega-constellation architecture.

\textbf{Operational cost model.} Operational costs are quantified through both storage and distribution energy consumption. Storage costs vary with node types and are modeled as follows:
\begin{equation}
PC_{cache}^{t} =\begin{cases}
\sum_{x\in GS} \sum_{f\in F}\beta_1\cdot B_f\cdot P_{x,f}^{t},  &x\in GS \\
\sum_{x\in GDC} \sum_{f\in F}\beta_2\cdot B_f,  & x\in GDC\\
\sum_{x\in S} \sum_{f\in F}\lambda_1\cdot B_f\cdot P_{x,f}^{t},    &x\in S \\
\sum_{x\in SDC} \sum_{f\in F}\lambda_2\cdot B_f,  & x\in SDC
\end{cases}\label{eq}
\end{equation}
where \(\beta_1\), \(\beta_2\),\(\lambda_1\), and \(\lambda_2\), represent unit power coefficients for ground stations, ground data centers, satellites, and space data centers, respectively. Distribution costs are calculated by
\begin{equation}
PC_{trans}^{t}=\begin{cases}
\displaystyle \sum_{f\in F}\sum_{x,y\in SN}\theta \cdot \delta^t(x,y)\cdot ISL_{(x,y)}^{t}\cdot B_f \\
\displaystyle \sum_{f\in F}\sum_{x\in GN, y\in SN}\eta  \cdot \delta^t(x,y)\cdot GSL_{(x,y)}^{t}\cdot B_f  \\
\displaystyle \sum_{f\in F}\sum_{x,y\in GN}\xi \cdot \delta^t(x,y)\cdot IGL_{(x,y)}^{t}\cdot B_f
\end{cases}  \label{eq}
\end{equation}
 incorporating link-specific transmission power parameters \(\theta\), \(\eta\), and \(\xi\), with \(\delta _{(x,y)}^{t}\in \left \{0,1\right \} \) indicating active content distribution between nodes \(x\) and \(y\). The coefficients $(\beta_1,\beta_2,\lambda_1,\lambda_2,\theta,\eta,\xi)$ are normalized per MB in $[0,1]$ (Table~\ref{tab1}). 

The total energy consumption over an operational period $T$ can be evaluated by
\begin{equation}
CT=\int_{0}^{T}PC_{cache}^{t} \mathrm{d}t+\int_{0}^{T}PC_{trans}^{t} \mathrm{d}t\label{eq}
\end{equation}
which aggregates time-varying storage and distribution costs across all tiers. This comprehensive model enables systematic optimization of the energy-latency trade-off while maintaining service quality guarantees for latency-sensitive satellite applications.

The primary objective is to minimize the total operational cost \(CT\) while satisfying the following constraints:
\begin{itemize}
\item Service accessibility is enforced by
\begin{equation}
\sum_{v\in V} \delta _{(u,v)}^t =1,\forall u \in U, v \in V,t\in T
\label{eq}
\end{equation}
ensuring all users \(u\in U\) connect to exactly one service anchor \(v\in V\) per time slot \(t\in T\). 
\item Content distribution feasibility is guaranteed through
\begin{equation}
\delta _{(x,y)}^t\le E_{(x,y)}^t,\forall x,y\in V, t\in T
\label{eq}
\end{equation}
which mandates that data transmission between nodes \(x\) and \(y\) only occurs when their interconnecting link \(E_{(x,y)}^t\) is active.
\item Latency requirements for content delivery are rigorously maintained via
\begin{equation}
\begin{split}
L_{(u,v_{cache}),f}^{t}\le \Delta _{upper\_latency}^{t} ,\\ 
\forall u\in U_{F,A}^{T} ,v_{cache}\in V,f\in F, t\in T
\end{split}
\label{eq}
\end{equation}
where the end-to-end delay \(L_{(u,v_{cache}),f}^{t}\) between user \(u\) and caching node \(v_{cache}\) must not exceed the predefined threshold\(\Delta _{upper\_latency}^{t}\). 
\item Storage capacity limitations are addressed by
\begin{equation}
\sum_{f\in F}P_{v,f}^{t}\cdot B_f \le C_v^t,\forall v\in V,f\in F, t\in T
\label{eq12}
\end{equation}
ensuring the cumulative size of cached content \(\sum_{f\in F}P_{v,f}^{t}\cdot B_f\) at any node \(v\) remains within its capacity \(C_v^t\).
\item System-wide redundancy control is implemented through
\begin{equation}
\begin{split}
\sum_{f\in F}\sum_{v\in V}P_{v,f}^{t}\cdot B_f \le \varphi \sum_{f\in F}B_f,\\
\forall v\in V,f\in F, t\in T
\end{split}
\label{eq13}
\end{equation}
which restricts the total cached content volume across all nodes to a fraction \(\varphi\) of the global content library size \(\sum_{f\in F}B_f\). This prevents excessive replication while maintaining service reliability.
\end{itemize}

\section{Algorithm Design and Performance Analysis}

\subsection{OSPF-based Satellite Path Computation (OSPC)}
OSPC (Algorithm~\ref{alg:OSPC}) adapts Dijkstra’s method to time-varying satellite graphs $S^t=(V,E^t)$ with dynamic link delays $L^t(x,y)$. It computes the minimum-latency path $SP^t$ from $v_{start}$ to $v_{end}$ in each snapshot.

\begin{algorithm}[!htbp]
\caption{OSPF-based Satellite Path Computation (OSPC)}
\label{alg:OSPC}
\textbf{Input:} $S^t=(V,E^t)$, $L^t(x,y)$, $v_{start}$, $v_{end}$ \\
\textbf{Output:} $SP^t$
\begin{algorithmic}[1]
\STATE Initialize $dist[v]\leftarrow\infty$, $prev[v]\leftarrow\emptyset$, $dist[v_{start}]\leftarrow 0$
\STATE Priority queue $Q\leftarrow V$ keyed by $dist[\cdot]$
\WHILE{$Q\neq\emptyset$}
  \STATE Extract $x\gets \arg\min_{v\in Q} dist[v]$
  \IF{$x=v_{end}$} \STATE \textbf{break} \ENDIF
  \FOR{each $(x,v)\in E^t$}
    \STATE $alt\leftarrow dist[x]+L^t(x,v)$
    \IF{$alt<dist[v]$} \STATE $dist[v]\leftarrow alt$; $prev[v]\leftarrow x$; \STATE Decrease-key in $Q$ \ENDIF
  \ENDFOR
\ENDWHILE
\STATE Reconstruct $SP^t$; \textbf{return} $SP^t$
\end{algorithmic}
\end{algorithm}

\textbf{Performance.}
For fixed $t$, OSPC returns the exact shortest-latency path. In dynamic scenarios with per-link delay drift $\le\epsilon_t$, the path stretch relative to the instantaneous optimum is bounded by $1+\frac{h\epsilon_t}{L^{t+1}(SP_*^{t+1})}$, where $h$ is the maximum hop count and $SP_*^{t+1}$ is the minimum-latency path in snapshot $S^{t+1}$. In multi-user settings with affine delay–load functions, per-flow OSPC achieves a constant-factor bound on total latency compared to the multi-commodity optimum. The per-recomputation complexity is $O((|E^t|+|V|)\log|V|)$.

\subsection{Multi-Level Collaborative Content Caching (MLC3)}
MLC3 coordinates hierarchical caching across three tiers:
\begin{itemize}
    \item \textbf{Tier-1}: Core data centers (GDC/SDC) store all high-value content permanently.
    \item \textbf{Tier-2}: Satellites cache content along OSPC paths when latency requirements are met, selected to minimize operational cost.
    \item \textbf{Tier-3}: Ground stations cache content for localized demand under latency constraints.
\end{itemize}

\begin{algorithm}[htbp]
\caption{Multi-Level Collaborative Content Caching (MLC3)}
\textbf{Input:} Connectivity graph $G^T=(V, E^T)$, content $F$, geographical region $A$, user request distribution $U_{F,A}^{T}$, delay requirement $\Delta_{\text{upper\_latency}}^{T}$ \\
\textbf{Output:} Content caching probability $P_{V,F}^{T}$, caching path $\delta^T(x,y)$, $\forall x \in S,y \in SDC$
\begin{algorithmic}[1]
\STATE Initialize: $P_{GS \cup S,F}^{T} \leftarrow 0$
\STATE \textit{/*MegaCacheX Tier-1 Cache*/}
\STATE $P_{GDC \cup SDC,F}^{T} \leftarrow 1$ 
\FOR{each $a \in A$}
    \FOR{each user request $u_{f,a}^{t}$ in increasing time $t$}
        \STATE \textit{/*MegaCacheX Tier-2 Cache*/}
        \STATE Acquire the user-access satellite $s_u$
        \STATE Obtain the satellite path $SP^t$ through Algorithm 1
        \FOR{satellite $s_k$ in path $SP^t$}
            \IF{$\max(L_{(u,s_{k}),f}^t) \le \Delta_{\text{upper\_latency}}^{t}$, $\forall u \in a_u$}
               \STATE $P_{s_k,f}^t,\delta^t(s_k,v)\leftarrow \arg\min_{f \in F} CT(U_{f,a}^t,s_k)$, $\forall v\in SDC$ 
            \ENDIF
        \ENDFOR
        \STATE \textit{/*MegaCacheX Tier-3 Cache*/}
        \FOR{each $gs \in GS$}
            \IF{$\max(L_{(u,v_{gs}),f}^t) \le \Delta_{\text{upper\_latency}}^{t}$, $\forall u \in a_u$}
                \STATE $P_{gs,f}^t \leftarrow \arg\min_{f \in F} CT(U_{f,a}^t,gs)$ 
            \ENDIF
        \ENDFOR
    \ENDFOR
\ENDFOR
\STATE \textbf{return} $P_{V,F}^{T}$, $\delta^T(x,y)$
\end{algorithmic}
\end{algorithm}

The algorithm iterates over all geographical regions \(a\in A\) and time slots \(t\in T\), dynamically adjusting caching probabilities \(P_{V,F}^{T}\) and distribution paths \(\delta^T(x,y)\) to balance storage utilization, latency compliance, and energy efficiency. This multi-tier coordination ensures proximity-based delivery while systematically controlling redundancy through \eqref{eq12}-\eqref{eq13}.

\textbf{Performance.}
MLC3 is a monotone submodular maximization under partitioned knapsack constraints (NP-hard). A greedy rule achieves an approximation of $(1-1/e)$ offline and $O(\log C_v)$ competitiveness online. Coupled with OSPC, the end-to-end bound is $(1-1/e)/\sigma$, where $\sigma={L^{t+1}(SP^{t})}/{L^{t+1}(SP_*^{t+1})}$, representing the worst-case multiplicative performance loss due to routing stretch when using previous-slot paths instead of the current-slot optimal paths. Per-update complexity is $O(|\mathcal{C}|\log|\mathcal{C}|)$, where $\mathcal{C}$ is the node–content candidate set restricted to $SP^t$ and its 1-hop neighbors; smaller $\mathcal{C}$ speeds execution at slight utility cost.

\section{PERFORMANCE EVALUATIONS}
The MegaCacheX testbed adopts a cloud-native microservices architecture, utilizing containerized infrastructure for mega-constellation routing and Kubernetes (K8s)-managed clusters for distributed caching. In this design, each container represents a physical node in the mega-constellation environment, implementing the satellite’s key capabilities for data storage and forwarding as required in our experiments. As the number of satellite nodes increases, a single server becomes incapable of hosting the large number of containers and lacks an effective resource scheduling mechanism. By leveraging K8s for container orchestration and lifecycle management, MegaCacheX overcomes challenges in cross-host communication and resource allocation, thereby enabling efficient scaling of the constellation size. Furthermore, the microservices-based design supports dynamic resource reallocation, automated failover, and load balancing, ensuring scalability and fault tolerance in large-scale, distributed deployments.
\begin{table*}[htbp]
\caption{Configuration and Cost Parameters of MegaCacheX}
\vspace{-12pt}
\begin{center}
\begin{tabular}{l c c}
\hline
\textbf{Component / Link} & \textbf{Configuration} & \textbf{Cost Parameter (per MB)} \\
\hline
Space Data Center (SDC) & 8 satellites, 1 orbit (700 km / 97.5$^\circ$) & $\lambda_2$ = 0.6 \\
Constellation Satellites (S) & 1584 satellites, 72 orbits (540 km / 53.2$^\circ$) & $\lambda_1$ = 1.0 \\
Ground Stations (GS) & 357 sites, distribution in Fig.~\ref{fig2}a & $\beta_1$ = 0.3 \\
Ground Data Centers (GDC) & 1285 nodes, distribution in Fig.~\ref{fig2}b & $\beta_2$ = 0.1 \\
User Request Regions & 1080 regions, distribution in Fig.~\ref{fig5}b &  \\
\hline
Inter-Satellite Link (ISL) & $+$Grid topology~\cite{hauri2020internet} & $\theta$ = 0.5 \\
Ground-to-Satellite Link (GSL) & Connect to the nearest (shortest-distance) satellite & $\eta$ = 1.0 \\
Inter-Ground Link (IGL) & Empirically measured data\cite{Globalping} & $\xi$ = 0.1 \\
\hline
\end{tabular}
\label{tab1}
\vspace{-18pt}
\end{center}
\end{table*}

\begin{figure*}[htbp]
    \center{\includegraphics[width=17cm] {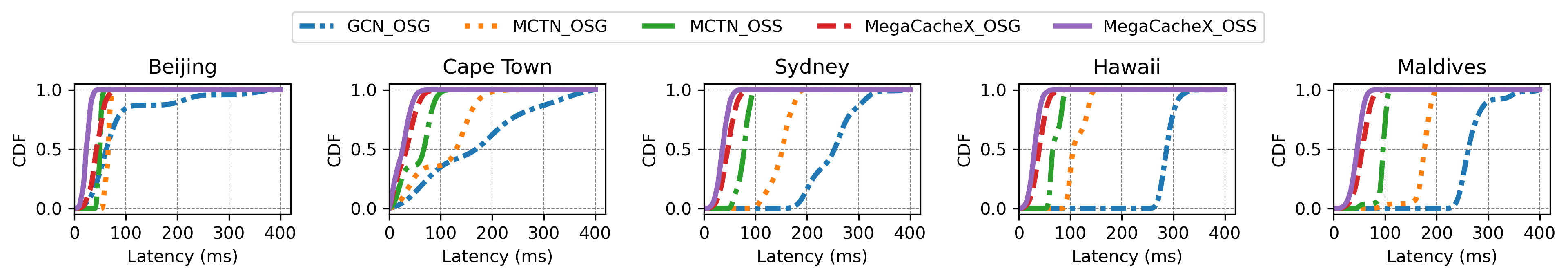}} 
    \vspace{-15pt}
    \caption{Global distribution of content response latency across different schemes.}
    \label{fig7}
    \vspace{-20pt}
\end{figure*}
MegaCacheX is instantiated in a hybrid space--ground environment (Table~\ref{tab1}). ISLs adopt a \(+\)Grid topology (each satellite connects to two same-orbit neighbors and two adjacent-orbit neighbors)~\cite{hauri2020internet}; GSLs associate with the nearest visible satellite; IGLs are characterized using empirically measured traces~\cite{Globalping}, with spatial distribution shown in Fig.~\ref{fig2}b. All cost parameters are normalized to \([0,1]\) to capture relative storage/transmission expenses across tiers: satellite storage and GSL transmission take the highest values (\(\lambda_1=1.0,\ \eta=1.0\)) due to tight power budgets and link dynamics; SDC storage and ISL forwarding are intermediate (\(\lambda_2=0.6,\ \theta=0.5\)); terrestrial resources are lowest, with GDC storage and IGL transport set to (\(\beta_2=0.1,\ \xi=0.1\)), and GS storage remaining modest (\(\beta_1=0.3\)). These settings encode operational asymmetries while enabling consistent, system-wide cost-aware optimization.Within this platform, Algorithm 1 and 2 are implemented as independent microservices and deployed in the K8s environment for scalable and modular execution.

We evaluate five network schemes comprising both ground-based and space-based origin server configurations: the Ground-based Cloud Network (GCN-OSG) with ground origin servers, the Mega-Constellation Transmission Network (MCTN-OSG and MCTN-OSS) with origin servers deployed either on the ground (OSG) or in space (OSS), and the Multi-level Collaborative Caching architectures (MegaCacheX-OSG and MegaCacheX-OSS), which integrate hierarchical caching mechanisms with identical origin server variants.
\begin{figure}[htbp]
\vspace{-12pt}
    \center{\includegraphics[width=\linewidth] {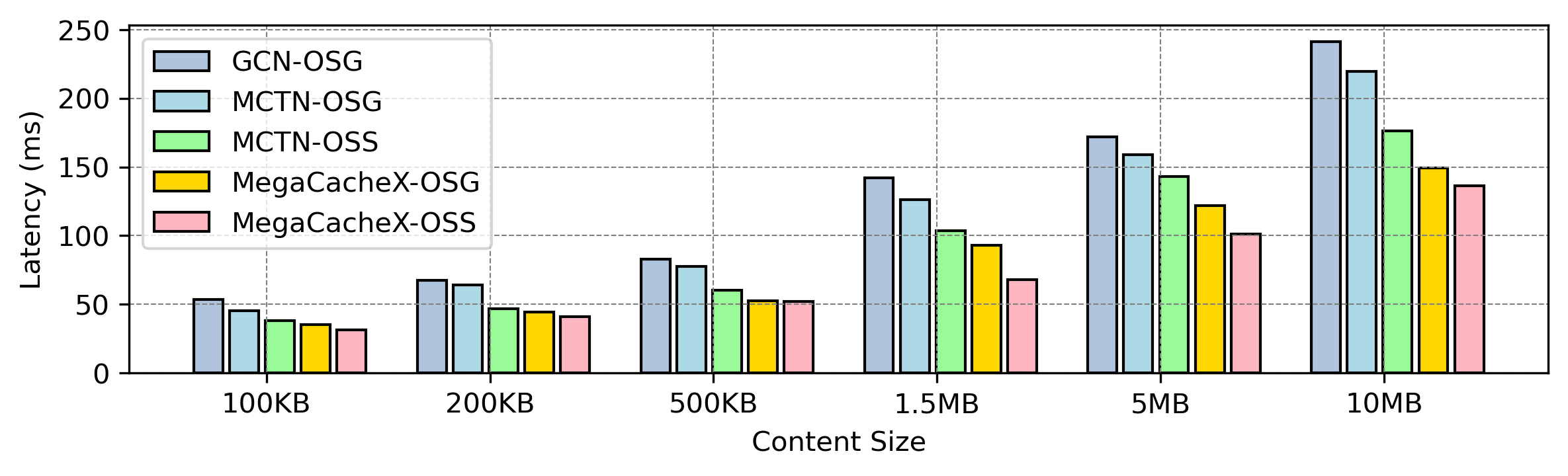}} 
    \vspace{-25pt}
    \caption{Latency characteristics under varying content sizes.}
    \vspace{-10pt}
    \label{fig8}
\end{figure}

As shown in Fig.~\ref{fig7}, MegaCacheX achieves optimal performance, guaranteeing sub-50 ms response times globally. Fig.~\ref{fig8} demonstrates a universal trend across all scenarios: smaller content sizes correlate with lower latency. Leveraging hierarchical caching, adaptive space-based routing, and optimized content placement, MegaCacheX reduces latency by 16\%–40\% compared to baselines.
\begin{figure}[htbp]
\vspace{-12pt}
    \center{\includegraphics[width=\linewidth] {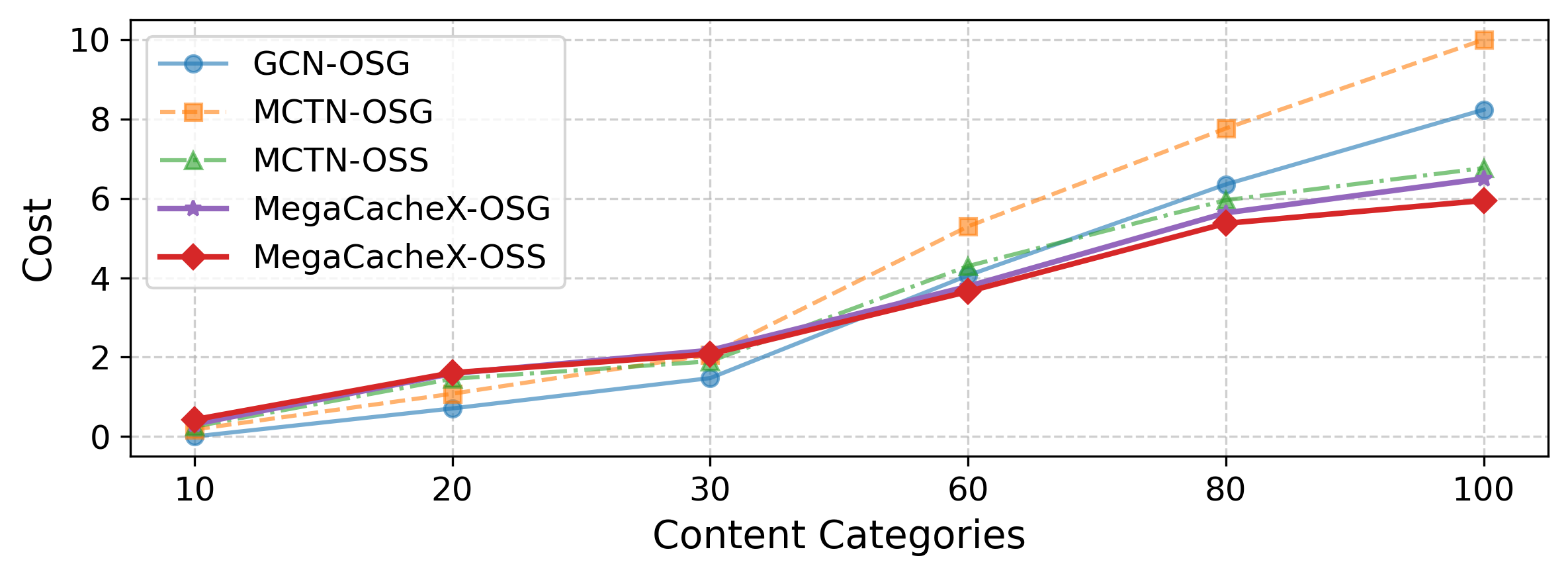}} 
    \vspace{-25pt}
    \caption{Cost characteristics of deployment schemes across content categories.}
    \vspace{-10pt}
    \label{fig9}
\end{figure}

In the analysis of the impact of different content types on scheme costs, the cost range of each scheme is scaled to 0-10 for easier visualization and discussion. Cost analysis in Fig.~\ref{fig9} reveals marginal differences among schemes when caching fewer than 30 content types, with MegaCacheX incurring slightly higher initial costs due to spaceborne storage and link expenses. However, as cached content volume increases, its three-tier architecture limits cost escalation, achieving up to 40\% cost reduction while maintaining competitive latency performance. This validates MegaCacheX’s efficacy in balancing operational economics with QoS guarantees.

\section{CONCLUSION}

MegaCacheX—a hierarchical caching framework for mega-constellations—integrates SSO data centers, adaptive intra-constellation caching, and ground-edge nodes to address the digital divide. By optimizing content placement based on real-time network conditions and demand, it reduces global access latency while maintaining cost-effectiveness, thereby overcoming the energy inefficiencies and redundancy limitations inherent in conventional systems. Empirical measurements confirm the necessity of orbital caching, while the modular design ensures scalability across heterogeneous constellations. This work provides a foundational blueprint for wide-area real-time communications, with future extensions aimed at AI-driven caching and in-orbit computing.
\section{Acknowledgment}
This work was supported by the National Project under Grant 2024-JCJQ-ZD-050-02 and the National Science Foundation of China under Grant 62271062.

\bibliographystyle{IEEEtran}
\bibliography{IEEEabrv,myrefs}

\vspace{12pt}

\end{document}